\documentstyle[12pt]{article}

\advance\textheight by \topskip
\begin{document}
\tolerance=100000
\thispagestyle{empty}
\setcounter{page}{0}

\newcommand{\mathrm}{\rm}
\newcommand{\ar}{\rightarrow}
\newcommand{\be}{\begin{equation}}
\newcommand{\ee}{\end{equation}}
\newcommand{\br}{\begin{eqnarray}}
\newcommand{\er}{\end{eqnarray}}
\newcommand{\ba}{\begin{array}}
\newcommand{\ea}{\end{array}}
\newcommand{\bi}{\begin{itemize}}
\newcommand{\ei}{\end{itemize}}
\newcommand{\bn}{\begin{enumerate}}
\newcommand{\en}{\end{enumerate}}
\newcommand{\bc}{\begin{center}}
\newcommand{\ec}{\end{center}}
\newcommand{\ul}{\underline}
\newcommand{\ol}{\overline}
\newcommand{\eettbbww}{$e^+e^-\rightarrow t^*\bar t^*\ar  b\bar b W^+ W^-$}
\newcommand{\phphttbbww}{$\gamma\gamma\rightarrow t^*\bar t^*\ar b\bar
b W^+ W^-$}
\newcommand{\eett}{$e^+e^-\rightarrow t\bar t\ar  b\bar b W^+ W^-$}
\newcommand{\phphtt}{$\gamma\gamma\rightarrow t\bar t \ar b\bar b W^+ W^-$}
\newcommand{\eebbww}{$e^+e^-\rightarrow b\bar b W^+ W^-$}
\newcommand{\phphbbww}{$\gamma\gamma\rightarrow b\bar b W^+ W^-$}
\newcommand{\bbww}{$b\bar b W^+ W^-$}
\newcommand{\uub}{$ u\bar u$}
\newcommand{\ddb}{$ d\bar d$}
\newcommand{\ssb}{$ s\bar s$}
\newcommand{\ccb}{$ c\bar c$}
\newcommand{\bbb}{$ b\bar b$}
\newcommand{\ttb}{$ t\bar t$}
\newcommand{\eeb}{$ e^+ e^-$}
\newcommand{\mumub}{$ \mu^+\mu^-$}
\newcommand{\tautaub}{$ \tau^+\tau^-$}
\newcommand{\veveb}{$ \nu_e\bar\nu_e$}
\newcommand{\vmvmb}{$ \nu_\mu\bar\nu_\mu $}
\newcommand{\vtvtb}{$ \nu_\tauu\bar\nu_\tau $}
\newcommand{\sm}{${\cal {SM}}$}
\newcommand{\MH}{$M_{H}$}
\newcommand{\Dir}{\kern -6.4pt\Big{/}}
\newcommand{\Dirin}{\kern -10.4pt\Big{/}\kern 4.4pt}
\newcommand{\DDir}{\kern -7.6pt\Big{/}}
\newcommand{\DGir}{\kern -6.0pt\Big{/}}
\def\Ord{\buildrel{\scriptscriptstyle <}\over{\scriptscriptstyle\sim}}
\def\OOrd{\buildrel{\scriptscriptstyle >}\over{\scriptscriptstyle\sim}}
\def\pl #1 #2 #3 {{\it Phys.~Lett.} {\bf#1} (#2) #3}
\def\np #1 #2 #3 {{\it Nucl.~Phys.} {\bf#1} (#2) #3}
\def\zp #1 #2 #3 {{\it Z.~Phys.} {\bf#1} (#2) #3}
\def\pr #1 #2 #3 {{\it Phys.~Rev.} {\bf#1} (#2) #3}
\def\prep #1 #2 #3 {{\it Phys.~Rep.} {\bf#1} (#2) #3}
\def\prl #1 #2 #3 {{\it Phys.~Rev.~Lett.} {\bf#1} (#2) #3}
\def\mpl #1 #2 #3 {{\it Mod.~Phys.~Lett.} {\bf#1} (#2) #3}
\def\rmp #1 #2 #3 {{\it Rev. Mod. Phys.} {\bf#1} (#2) #3}
\def\xx #1 #2 #3 {{\bf#1}, (#2) #3}
\def\preprint{{\it preprint}}

\begin{flushright}
{\large DFTT 39/95}\\
{\large DTP/95/56}\\
{\rm June 1995\hspace*{.5 truecm}}\\
\end{flushright}

\vspace*{\fill}

\begin{center}
{\Large \bf Finite width and irreducible background effects to
$t\bar t$ production at $\gamma\gamma$ Next Linear
Colliders\footnote{Work
supported in part by Ministero
dell' Universit\`a e della Ricerca Scientifica.\\[4. mm]
E-mails: Moretti@to.infn.it;
Stefano.Moretti@durham.ac.uk.}}\\[2.cm]
{\large Stefano Moretti\footnote{Address
after September 1995: Cavendish Laboratory,
University of Cambridge,
Madingley Road,
Cambridge, CB3 0HE, U.K.}}\\[0.5 cm]
{\it Dipartimento di Fisica Teorica, Universit\`a di Torino,}\\
{\it and I.N.F.N., Sezione di Torino,}\\
{\it Via Pietro Giuria 1, 10125 Torino, Italy.}\\[0.5cm]
{\it Department of Physics, University of Durham,}\\
{\it South Road, Durham DH1 3LE, United Kingdom.}\\[0.75cm]
\end{center}
\centerline{PACS number(s): 13.65.+i, 14.70.Bh, 13.60.Fz, 14.65.Ha.}
\vspace*{\fill}

\begin{abstract}
{\normalsize
\noindent
We study the complete process \phphbbww\ using exact matrix element
computations at tree-level, at a $\sqrt s=500$ GeV
\eeb\ linear collider of the next generation. Incoming
photons produced via back-scattering of laser light are considered.
Sizable effects due to the finite width of the top quark
as well as to the irreducible background to $t\bar t$ production
and decay are predicted.}
\end{abstract}

\vspace*{\fill}
\newpage
\subsection*{1. Introduction}

In the past few years the importance of studying the
\ttb\ threshold region through $\gamma\gamma$ collisions at an \eeb\
Next Linear Collider (NLC) has clearly come out.

The option of high energy photon-photon interactions at a NLC
would principally make use of well focused high energy $\gamma$'s produced
through back-scattering of laser light from the electron/positron
beams \cite{back}, even though photonic interactions can take place
also via beamsstrahlung \cite{beam}, as well as via conventional
bremsstrahlung.
Photons can interact through their quark and gluon
constituents (`resolved' photons). However, in the case of
\ttb\ production the resolved photon rates are fairly small: this
mainly depends on the steeply falling parton distributions inside the
photon itself\footnote{In our analysis we deal with
back-scattered photons and with
`direct' $\gamma\gamma$ interactions only.}.

It has been established by recent measurements that
$m_t\OOrd 160$ GeV \cite{top}, therefore the top lifetime is expected to
be shorter than the typical time-scale of strong interactions (i.e.,
$\Gamma_t>\Lambda_{QCD}$). In this case, within
the Standard Model (\sm), top quarks
should decay to $bW$ pairs before toponium formation can occur.
This has two important consequences. On the one hand, we do not expect
any narrow QCD resonance around the energy $2m_t$ (as for \ccb\ and
\bbb\ bound states) and, on the other hand, the
large top width ($\Gamma_t\OOrd 1.5$ GeV) introduces a natural infrared
cut-off, such that it is possible to study all the threshold region
within the context of perturbation theory \cite{FK,BDKKZ}. In fact, the
strong running coupling constant $\alpha_s$ occurs at the scale
$Q^2\approx m_t\sqrt{\Gamma_t^2+E^2}$ (with $E=\sqrt s - 2m_t$), which
never gets small.

The main possibilities offered by the $\gamma\gamma\ar t\bar t$
process are the following. First, by measuring the shape and the
position of the \ttb\ threshold one could determine both $m_t$ and
$\alpha_s$ \cite{threshold1,threshold2}.
Second, one can measure
$\Gamma_t$ (thereby checking the value of the
Cabibbo-Kobayashi-Maskawa term $|V_{tb}|$)
if an energy resolution of $\Delta E_{\gamma\gamma}\Ord 1$
GeV can be achieved \cite{Chanowitz}. Third, with $\OOrd 95$\%
polarized $\gamma$ beams of opposite helicities (which strongly
suppress the dominant $s$-wave) it is possible to directly study
\ttb-production in $p$-wave, whereas
for \eeb\ collisions $p$-wave production can be probed only
by its interference with the $s$-wave one \cite{threshold1,wave}.
Finally, by linearly polarized photon beams one is able to
determine the top quark polarization, thus providing a powerful tool to
study final QCD interactions and to precisely measure
$\alpha_s$ \cite{linearly}. The cross section for
$\gamma\gamma\ar t\bar t$ is comparable to the one of
$e^+e^-\ar t\bar t$, becoming
even larger for $\sqrt s\OOrd 500$ GeV.

It is the aim of this brief report to add further elements to the argument,
by computing all the irreducible background in $\gamma\gamma\ar b\bar
bW^+W^-$ events, as well as by treating finite top width effects in the
appropriate way, without resorting neither to the off-shell
production$\times$Breit-Wigner decay approach nor to
the Narrow Width Approximation (NWA),
while keeping into account all spin and energy correlations between
the two top decays. We do not include here the QCD
(and QED)
Coulomb-like interactions between the two top quarks at threshold
(they lead to an enhancement of the cross section).
We do not do that for two reasons. First, the large top
mass allow us to conclude that the resonant structures in \ttb\
production are largely
smeared out and do not show up dramatically in the excitation curve.
Second, since the Coulomb corrections to \ttb\
production are  part of the higher order corrections
and the non resonant background is evaluated here at tree-level,
 we have ignored them for consistency.
For the same reason we have not computed gluon
bremsstrahlung radiative corrections.

This study resembles the corresponding analysis
done for $e^+e^- \ar b\bar b W^+W^-$ in ref.~\cite{eett}.
We turned to the $\gamma\gamma$ case since we expect the
finite width effects of the top to be largely independent of the
production mechanism, therefore they should remain important in this
context as well. In addition, since the process
$\gamma\gamma\ar b\bar bW^+W^-$ involves 114 Feynman diagrams at
tree-level, compared to the 61 occurring in $e^+e^- \ar b\bar b
W^+W^-$, we expect the irreducible background to be quantitatively even more
important here.
Finally, the fact that the incoming photons do not have a fixed
energy and momentum, but these are spread according to the
back-scattering spectrum, introduces a smearing in the
differential distributions, which tends to wash out effects typical
of \ttb\ resonant production.
Similar effects occur also in connection with, e.g., the Initial State
Radiation (ISR) in \eeb\ annihilations. We expect the signal to be
sensitive to all of these more than the non resonant
$\gamma\gamma\ar b\bar bW^+W^-$ production.

The paper is structured as follows. In sect.~2 we give details of
the calculations. In sect.~3 we describe and
comment on the results we obtained. A brief summary including
our conclusions will
be given in sect.~4.

\subsection*{2. Computation}

The process \phphbbww\
is described at tree-level by 114 Feynman graphs.
Their structure is very rich,
including (apart from Higgs self interactions) all the types of \sm\ couplings
at tree-level\footnote{Therefore they also constitute a useful tool to
test the model itself.}.
Seven subsets of the graphs are displayed in fig.~1. Out of the original
114 diagrams, we tried to separate the most topologically different
ones. We recognised the following structures, of diagrams involving:
a) four vertices $Vff'$, where $V=\gamma,Z$ or $W$,
         and $ff'=bb$ or $bt$: 12 graphs;
b) three vertices $Vff'$, and one $W^+W^-V'$, where
          $V=\gamma,Z$, or $W$, $V'=\gamma$ or $Z$,
          and $ff'=bb$ or $bt$: 24 graphs;
c) two vertices $Vff'$, and two connected $W^+W^-V'$, where
          $V=\gamma,Z$, or $W$, $V'=\gamma$ or $Z$,
          and $ff'=bb$ or $bt$: 20 graphs;
d) two vertices $Vff'$, and two disconnected $W^+W^-V'$, where
          $V=W$, $V'=\gamma$,
          and $ff'=bt$: 2 graphs;
e) one vertex $Vff'$, and three (connected) $W^+W^-V'$, where
          $V=V'=\gamma$ or $Z$, and $ff'=bb$: 12 graphs;
f) two vertices $Vff'$, and one $W^+W^-\gamma V'$, where
          $V=\gamma,Z$, or $W$, $V'=\gamma$ or $Z$,
          and $ff'=bb$ or $bt$: 10 graphs;
g) one vertex $Vff'$, one $W^+W^-\gamma$, and one $W^+W^-\gamma
          V'$, where $V=V'=\gamma$ or $Z$,
          and $ff'=bb$: 12 graphs.

All the necessary diagrams can be obtained by the topologies of
fig.~1 by properly labeling both the internal and external
lines following the indications given in a--g. In addition, to get the
contributions involving internal Higgs lines (22 graphs) one has
to replace for
any virtual $V=\gamma,Z$ the $H$, apart from the case of quartic couplings
$W^+W^-\gamma V$. The different sets in a--g correspond to
different structures of the {\tt FORTRAN} routines too.

The matrix element for \phphbbww\ has been computed with the
help of MadGraph/HELAS \cite{Tim}. In order to keep the
interplay between the various resonances
which appear in the integration domains of the final state \bbww,
when all tree-level contributions are kept into account, under
control,
we have adopted the technique of splitting the Feynman amplitude squared into
a sum of different (non-gauge-invariant) terms, each of which has been
integrated according to its resonant structure with an appropriate
choice of integration variables and phase space. The gauge invariance
is recovered by summing at the end
the integrated contributions
of the above terms, both in the total and the differential rates.
This procedure has been carefully described elsewhere \cite{eezh},
so we do not enter here into details.
The multi-dimensional integrations over the phase space have been
performed numerically using VEGAS \cite{Vegas}.\par

The following  values of the parameters have been adopted:
$M_{Z}=91.1$ GeV, $\Gamma_{Z}=2.5$ GeV,
$M_{W}\equiv M_{Z}\cos(\theta_W)\approx80$ GeV,
$\Gamma_{W}=2.2$ GeV, and $\sin^2 (\theta_W)=0.23$.
For the fermions: $m_b=5$ GeV and $m_t=175$ and 200 GeV, according to
the values announced by CDF and D0. The top width $\Gamma_t$
has been computed at tree-level, for coherence\footnote{This is needed
in fact to recover the $\gamma\gamma\ar t\bar t$ cross section in NWA
(see later on).}.
In order to get rid of complications due to virtual Higgs contributions
we have deliberately
decided to set $M_{H}$ equal to a value which
strongly suppresses them (at $\sqrt s=500$ GeV, e.g., $M_{H}= 700$ GeV).
Also for the Higgs width we have adopted the expression at tree-level.
The electromagnetic coupling constant has been set equal to 1/128.
We have carried out our analysis at the standard energy $\sqrt
s=500$ GeV.
Finally,
we have not implemented the effects due to the finite width of the final state
$W$'s: we are confident, however,  that taking them into account would
not affect our conclusions.

\subsection*{3. Results}

The results we obtained are presented throughout tab.~I and
figs.~2--4. In the following we will adopt the notations `NWA',
`Production \& decay', `All diagrams', to indicate the three processes
\phphtt, \phphttbbww and \phphbbww, respectively. In the first reaction, the
top quarks are produced on-shell and subsequently decay to $bW$
pairs. In order to achieve this we have kept into the computations only the
two graphs of `type a' in which both the photons couple to the top
line, and we have re-written the top propagator as
\be\label{prop}
\frac{p\Dir +m_t}{p^2-m_t^2+{\rm{i}}m_t\Gamma}
\left(\frac{\Gamma}{\Gamma_{{t}}}\right)^{1/2},
\ee
then we have taken $\Gamma\ar 0$. In this limit, the square of
eq.~(\ref{prop}) produces a $\delta(p^2-m_t^2)$ (i.e., giving on-shell
\ttb\ production). Numerically, we used $\Gamma=10^{-5}$, which
produces results in very good agreement with those from
 the two-to-two body
reaction $\gamma\gamma\ar t\bar t$.
In the second process, the top quarks are produced also off-shell by setting
$\Gamma=\Gamma_{{t}}$ with $\Gamma_t$ finite (so the standard
expression from the propagator is recovered), and by using again the two
above diagrams only. In the third case, all the diagrams entering at
tree-level into the process \phphbbww\ are computed (graphs a+b+c+d+e+f+g),
with the same convention as in \phphttbbww\ for the top propagators
(apart from the case of $t$-channel propagators, which have no
imaginary part).

Tab.~I shows that the irreducible background to $t^{(*)}\bar
t^{(*)}$ production and decay is not negligible, since
it gives an additional contribution which is $\approx 21(500)\%$
for $m_t=175(200)$ GeV, with respect the on-shell
\ttb\ production: much(enormously) larger than in the case of
\eeb\ initiated top pair production (see ref.~\cite{eett}).
The difference between the case `NWA' and `Production \& decay' are
of $\approx3(20)\%$, and the rates are larger in the first(second) case.
This reflects the fact that, for $m_t=175$ GeV,
increasing the top width in the square
of the propagator (\ref{prop}) reduces the total cross section more
than phase space reduction by requiring two on-shell top quarks (see also
tab.~II in ref.~\cite{eett}), whereas if $m_t=200$
GeV the situation is the other way round.

In order to understand the large difference in the total cross sections
for the two different values of $m_t$, we notice that
the peak in the energy distribution of a backscattered
photon is roughly at $\approx0.8\times(\sqrt s/2)\approx 200$ GeV.
The fact that the cross section for $m_t=200$ GeV is much smaller than
the one for $m_t=175$ GeV indicates that the quantity
$E=\sqrt s -2m_t$ is negative most
of the times (i.e., \ttb\ production below threshold occurs).

Since we expect $\Gamma_t$ to influence the kinematics of the top decay
products $b$ and $W$, we studied in fig.~2 the dependence of the cross
sections, e.g., on the momentum of
the $W$, for \phphtt,
\phphttbbww\ and \phphbbww. Whereas, in the case of \eeb\
annihilation, the fact that the system \eeb\ has  a fixed energy
(apart from beamsstrahlung and ISR effects),
equal to $\sqrt s$,
allows to opportunely tune the collider energy
in order to study \ttb\ production at
few GeV above threshold and to deduce the
top mass from the $p_W$ spectrum \cite{Bagliesi}, in photon-photon
collision this is quite problematic. In fact, the invariant mass of the
$\gamma\gamma$ system is not fixed but follows a luminosity
distribution. Therefore, no clear edge appears in the $p_W$
distribution in the case of NWA. Also, contrary to the
case of $e^+e^-$ top production, no systematic difference in the shape
of the curves for \phphtt, \phphttbbww and \phphbbww\
occurs (see fig.~2 in ref.~\cite{eett}).

However, this variable represents a good choice for studying the
effects due to the increase of the top width, connected with
the possible existence of new physics beyond the \sm.
In fact, fig.~3 shows the
strong sensitivity of $p_W$ on $\Gamma_t$, in the case this latter
is increased by a factor of 1.25 and 1.5. The integrated cross
sections give 50.16(9.32) and 34.83(8.12) fb, respectively, for the two
usual values of the top mass, to be compared with the values in the third
column of tab.~I. Here, the complete process \phphbbww\ is considered.

In order to get rid of the irreducible background,
a natural procedure appears to be cutting
around the narrow top peak in the invariant mass of the $bW$ system
($\Gamma_t\Ord 2.5$ GeV for
$m_t\Ord200$ GeV).
However, once one has reconstructed the $W$ from its decay products
(leptons or hadrons)
there is still an ambiguity in assigning the third
jet, since this can be produced either by the `right' $b$ (the one coming
in association with the tagged $W$ in the same top decay) or by
the `wrong' $b$ (the one coming from the other top decay). Therefore,
one usually constructs two combinations $bW$, with only one peaking at $m_t$.
A way of avoiding this could be to recognise the charge of the parton
from which the jets originate (e.g., by the `jet charge' method or
by tagging the lepton from the decaying $b$). However, this is
unlikely to
give high efficiencies, since in order to enhance the signal rates the
other $W$ is usually tagged by its hadronic decays, therefore, one ends up
dealing with at least 4 jets in the final state.

In fig.~4 we show the invariant mass of the right and wrong $bW$
combinations, plotted in 10 GeV bins. For example, the areas in the range
$|M_{bW}-m_t|\le10$ GeV are $\approx 60(4.81)$ and  $\approx 6(0.88)$ fb,
for $m_t=175(200)$ GeV, respectively.
Therefore the contamination of unwanted wrong
$bW$ combinations in peaking up events around the top
mass is $\approx10(18)\%$, depending on the top mass itself.
These percentages roughly represent the size of the errors
in the analyses which select the candidate \ttb\ sample by cutting in the
reconstructed $bW$  invariant mass.
At the same time, the two
distributions in fig.~3 sum up to just one, with the effect of
making the top peak broader and the
eventual determination of $\Gamma_t$ from this spectrum
quite problematic. This effect
is more important here than in \eeb\ collisions.

\subsection*{4. Summary and conclusions}

We have studied differential and integrated rates for the processes
\phphtt, \phphttbbww\ and \phphbbww, by incoming photons generated
via Compton back-scattering of laser light. We used exact matrix element
calculations at tree-level. In the first case only the on-shell
$t\bar t$-production is computed, in the second reaction
also off-shell and finite width effects of the top are included,
whereas in the third process all the gauge invariant
set of Feynman diagrams is considered and no approximation is adopted.
This allowed us to estimate that the irreducible background
in \bbww\ final states to the process $\gamma\gamma\ar t\bar t$
at a NLC with $\sqrt s=500$ GeV
increases the integrated
\ttb\ signal rates by $\approx21\%$(a factor of 5), if
$m_t=175(200)$ GeV. At $\sqrt s=500$ GeV,
the two values $m_t=175$ and $200$ GeV
correspond to the two opposite cases in which $t\bar t$-pairs
are produced above and below the threshold $2m_t$, respectively. As
for a NLC operating in the $\gamma\gamma$-mode the Center-of-Mass
(CM) energy of the photons is not fixed but follows a luminosity
distribution, both of these are realistic conditions.
Therefore, this clearly shows
how both an excellent determination of $m_t$
and a careful tuning of the energy $\sqrt s$ of the $e^+e^-$ system
are needed, in order to control non resonant \bbww\ events and to
study in detail the $t\bar t$--threshold via
$\gamma\gamma$-collisions.\par
The top finite width effects have been studied by comparing the
on-shell \ttb\ production and decay in NWA with the off-shell one.
Here, differences vary from 3 to 20\%, depending on the top mass.
Moreover, if the top width turns out to be larger than the \sm\ one,
due to possible new physics, sizable effects are expected.
\par
Before drawing definite conclusions these results should be folded
with a realistic simulation including the expected performances
of the detectors of a NLC, and studied depending on the adopted
experimental strategies. High order corrections (Coulomb
singularities, gluon and photon radiative corrections, ISR, etc ...)
should be properly included as well.
However, our predictions seem to indicate that
effects due to the unavoidable presence of the irreducible background
and to the finite width of the top
should be included in the phenomenological analyses.

\subsection*{Acknowledgements}

We are grateful to V.A.~Khoze for useful comments and for reading the
manuscript.

\vfill
\newpage
\thispagestyle{empty}

\subsection*{Table Captions}

\begin{description}

\item[table~I] Cross sections for the processes \phphtt\ (NWA), \phphttbbww\
(Production \& decay) and \phphbbww\ (All diagrams),
at $\sqrt s=500$ GeV, for the two values of the top
mass $m_t=175$ and 200 GeV. For the Higgs mass we have taken $M_H=700$
GeV.

\end{description}
\
\vskip1.0cm

\subsection*{Figure Captions}

\begin{description}

\item[figure~1 ] The topologically different subsets of Feynman diagrams
contributing at tree-level to \phphbbww (see the text).
Continuous lines represent a $b$ or a $t$, whereas wavy lines refer
to a $\gamma$, a $W$, a $Z$ or a $H$, as appropriate, according
to the couplings within the \sm.

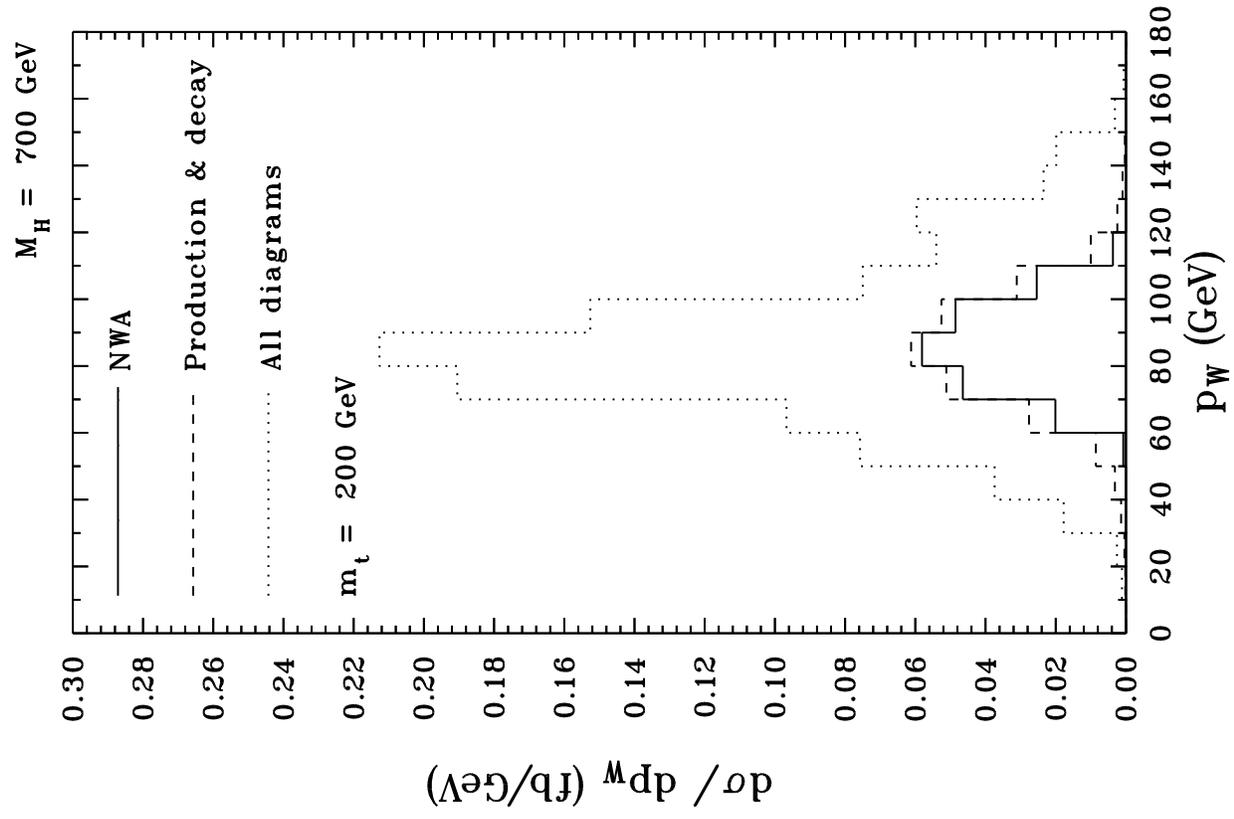
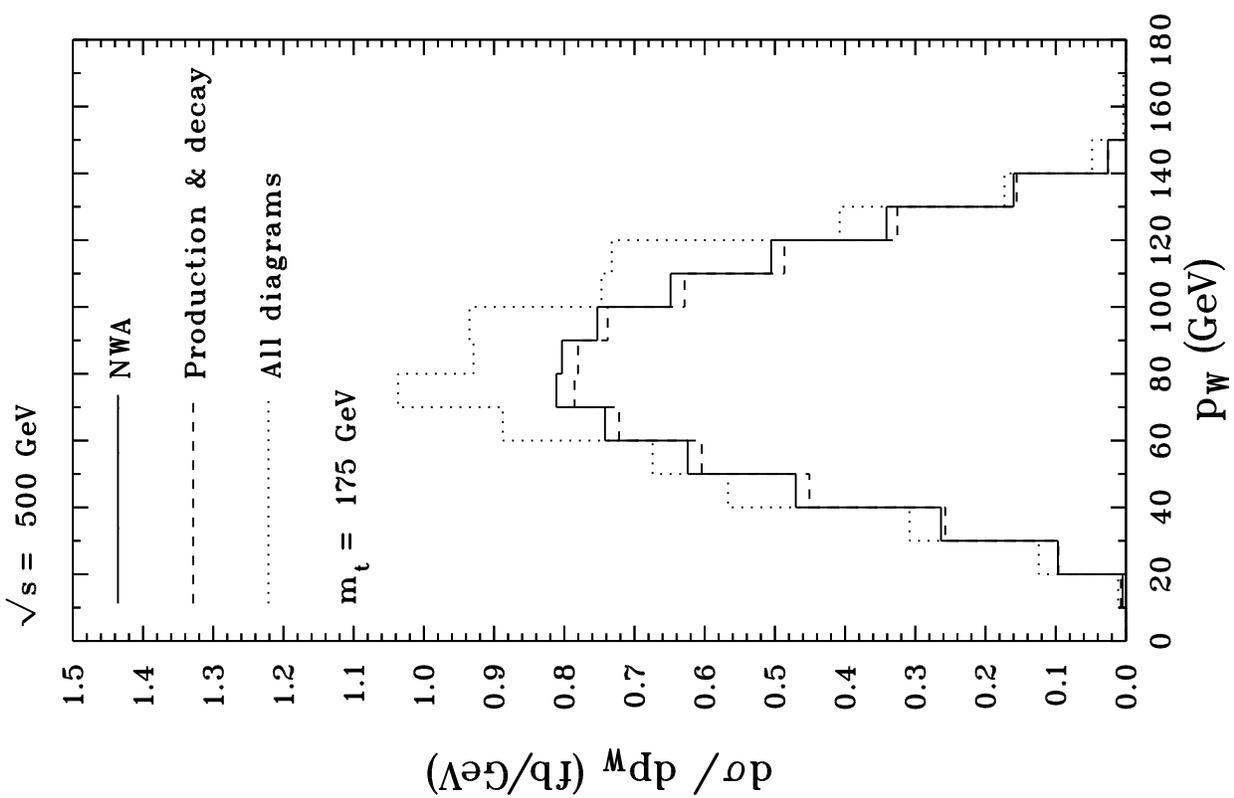
\item[figure~2 ] Differential distribution in the momentum of
the $W$, for the three cases \phphtt\ (continuous lines),
\phphttbbww\ (dashed lines) and \phphbbww\ (dotted lines),
at $\sqrt s=500$ GeV, for the two values of the top
mass $m_t=175$  and 200  GeV. For the Higgs mass we have taken $M_H=700$
GeV.

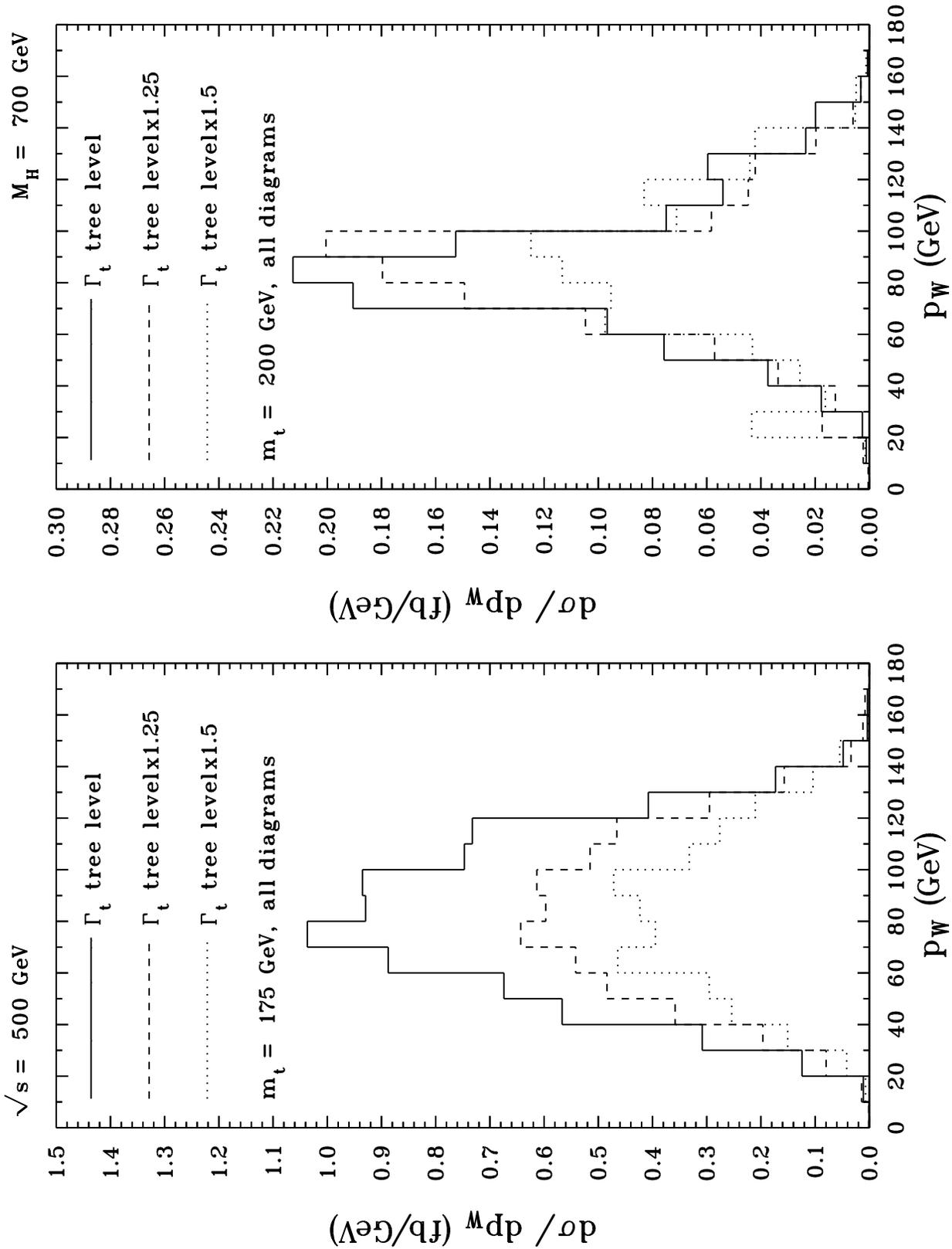
\item[figure~3 ] Differential distribution in the momentum of
the $W$, in the case \phphbbww\ only, for the \sm\ tree-level
top width $\Gamma_t$ (continuous lines), for $1.25\times\Gamma_t$
(dashed lines) and $1.5\times\Gamma_t$ (dotted lines),
at $\sqrt s=500$ GeV, for the two values of the top
mass $m_t=175$ and 200 GeV. For the Higgs mass we have taken $M_H=700$
GeV.

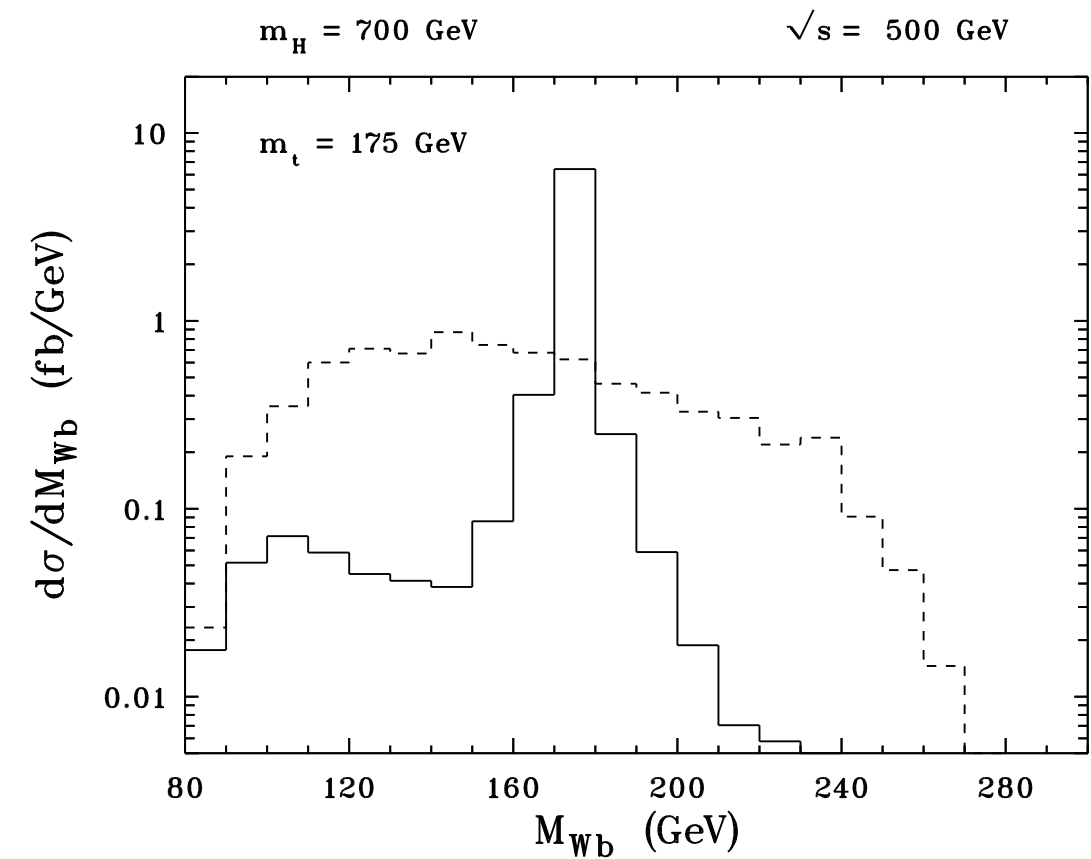
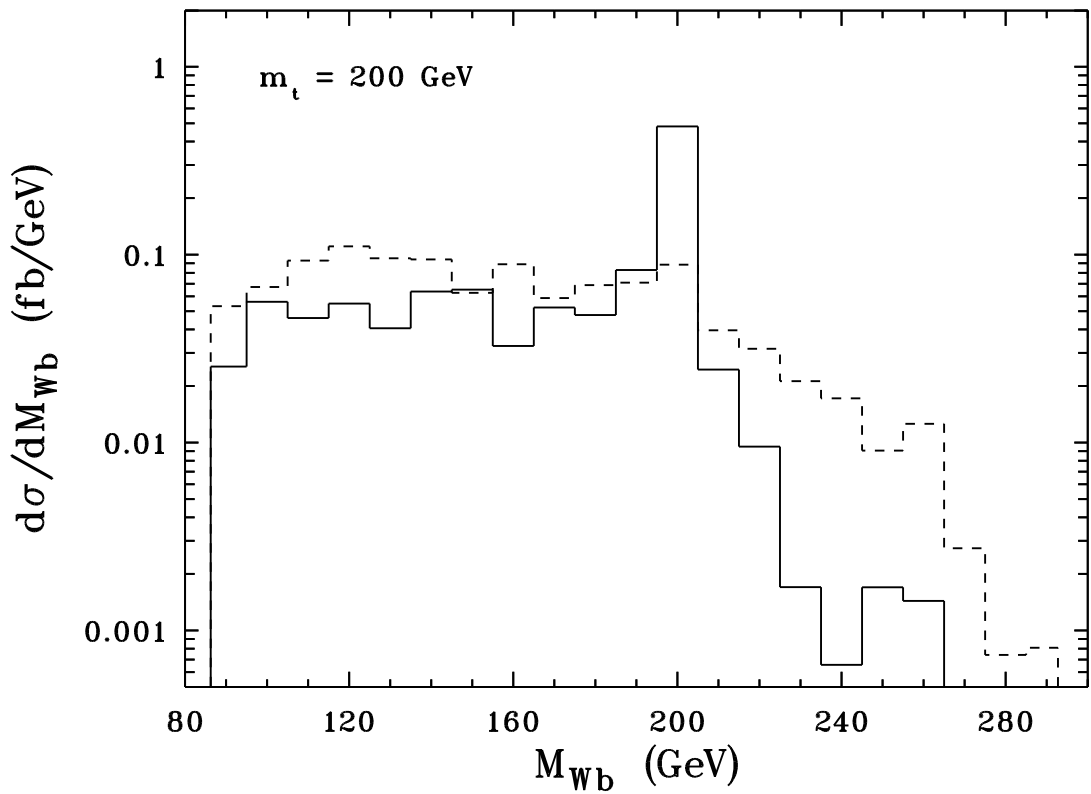
\item[figure~4 ] Differential distribution in the invariant mass of
the $bW$ system, in the case \phphbbww, for the `right' (continuous
lines) and the `wrong' (dashed lines) combinations of $b$ and $W$ (see
the text), at $\sqrt s=500$ GeV, for the two values of the top
mass $m_t=175$ and 200 GeV. For the Higgs mass we have taken $M_H=700$
GeV.

\end{description}

\vfill
\newpage
\thispagestyle{empty}
\begin{table}
\begin{center}
\begin{tabular}{|c|c|c|c|}
\hline
\multicolumn{4}{|c|}
{\rule[-0.5cm]{0cm}{1.1cm}
$\sigma(\gamma\gamma\ar X)~{\rm{(fb)}}$}
 \\ \hline
\rule[-0.5cm]{0cm}{1.1cm}
   $m_t~{\rm{(GeV)}}$
&  {\rm{NWA}}
&  {\rm{Production~\&~decay}}
&  {\rm{All~diagrams}}  \\ \hline\hline
\multicolumn{4}{|c|}
{\rule[-0.5cm]{0cm}{1.1cm}
 $~~\sqrt s=500~{\rm{GeV}}$ }
\\ \hline
\rule[-0.6cm]{0cm}{1.3cm}
$175$ & $62.49$ & $60.64$ & $75.90$    \\ 
\rule[-0.6cm]{0cm}{.9cm}
$200$ & $2.03$ & $2.51$ & $10.22$     \\  \hline\hline
\multicolumn{4}{|c|}
{\rule[-0.5cm]{0cm}{1.1cm}
$M_H=700$~GeV}
 \\ \hline
\multicolumn{4}{c}
{\rule{0cm}{.9cm}
{\Large Table I}}  \\
\multicolumn{4}{c}
{\rule{0cm}{.9cm}}

\end{tabular}
\end{center}
\end{table}

\vfill

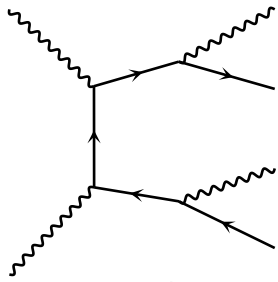
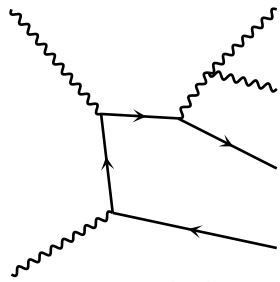
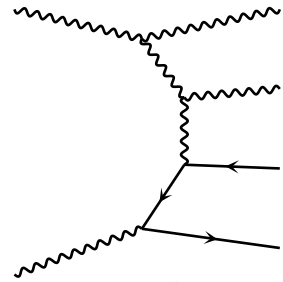
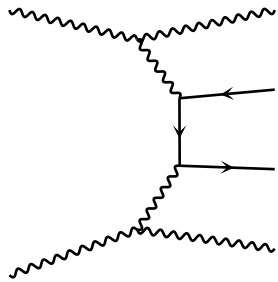
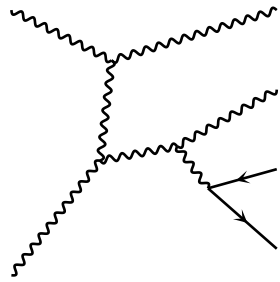
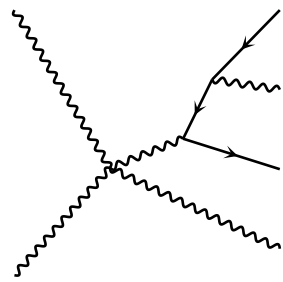
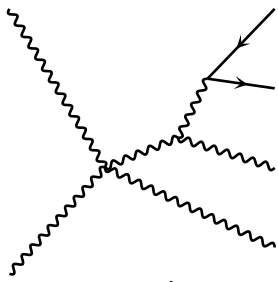
\end{document}